\documentstyle[12pt,aasms4]{article}

\def\gappr{\mathrel{\vcenter{\offinterlineskip \hbox{$>$}
    \kern 0.3ex \hbox{$\sim$}}}}
\def\lappr{\mathrel{\vcenter{\offinterlineskip \hbox{$<$}
    \kern 0.3ex \hbox{$\sim$}}}}

\def\fun#1#2{\lower3.6pt\vbox{\baselineskip0pt\lineskip.9pt
        \ialign{$\mathsurround=0pt#1\hfill##\hfil$\crcr#2\crcr\sim\crcr}}}

\def\ben{\begin{equation}}
\def\be#1{\begin{equation}\label{eq:#1}}
\def\ee{\end{equation}}

\received{}
\revised{}
\singlespace

\begin{document}

\title{On the significance of the observed clustering of 
ultra-high energy cosmic rays}

\author{Gustavo A. Medina Tanco$^{1,2}$}

\affil{ 1.
Instituto Astron\^omico e Geof\'{\i}sico, University of S\~ao Paulo, Brasil \\
gustavo@adromeda.iagusp.usp.br
}

\affil{ 2. Royal Greenwich Observatory, Cambridge, UK}

\singlespace

\begin{abstract} 

Three pairs of possibly correlated ultra-high energy cosmic ray events were 
reported by Hayashida et al (1996). 
Three different numerical models are combined to study the propagation
of the corresponding particles through both the intergalactic and galactic 
magnetic fields. The spatial dependences of fields and galaxies are accounted 
for. The results suggests that the pairs are chance clusterings.




\end{abstract}

\clearpage
                              
\section{Introduction}

Recently, Hayashida et al. (1996) reported the possible clustering of 
some of the ultra high energy events of the AGASA experiment.
If these ultra high 
energy cosmic rays (UHECR) are charged particles, protons as it is more
likely, then these pairs impose severe constraints on the 
characteristics of the propagation region and/or their sources 
(e.g., Cronin 1996, Sigl et al 1996). Catastrophic extragalactic 
events, like $\gamma$-ray bursts (GRB) or the decay of topological defects 
(TD), which are able to produce the particles over a very short period 
of time, should only be consistent with the data for a suitable 
combination of  low intergalactic magnetic field (IGMF) and distance 
to the source. Nevertheless, the stirring of the intergalactic 
medium (IGM) by large agglomerates of galaxies, shocks excited 
in binary collisions of galaxies or the bow shocks preceding fast 
moving galaxies in dense IGM environments are examples of quiescent 
sources that could produce chance pairings of UHECR events on the 
sky. If these quiescent sources are traced by the distribution of 
luminous matter in the nearby universe, which is known, then the 
probability of the corresponding chance pairing can be estimated 
and compared with the observations.  

In this letter, the results of three different calculations are presented. 
First, the trajectories of the individual particles through the galactic 
magnetic field (GMF) are calculated for each pair under different 
assumptions for the GMF (Medina Tanco et al, 1997a, Medina Tanco 1997a). 
In the case of catastrophic events (i.e, almost simultaneous particle 
emission) this constrains the amount of time delay due to intergalactic 
propagation alone and, consequently, the range of IGMF values and source
distances allowed. The separation angle between the momenta of the 
particles at their arrival at the border of the halo, $\theta_{HALO}$,
can also be estimated. This is a matching condition that must be 
satisfied by the particle trajectories at the border of the halo. 
Second, the same numerical scheme of  Medina Tanco et al (1997b) is 
used to estimate the arrival relative-deflection distribution function 
for some allowed combinations of IGMF and distance to the source. 
The comparison of this distribution function with the previously 
calculated $\theta_{HALO}$, gives a quantitative idea of the 
likelihood of the observed events being the result of point-like 
catastrophic sources. Third, the actual distribution of extragalactic 
objects, as given by the CfA catalogue (Huchra et al 1995), is 
assumed to track the UHECR sources and to modulate the intensity 
of the IGMF. Consequently, with the aid of numerical three dimensional 
simulations, an all-sky arrival probability distribution function of 
UHECR is built (Medina Tanco 1997b,c) and compared with the observations.

\section{Numerical models and discussion of results}

Three different codes are used in the present work. The first one allows 
the calculation of the trajectory of an UHECR particle of known mass 
and charge between the border of the galactic halo and the detector 
at Earth. A complete description of the model can be found in Medina 
Tanco et al (1997a) and Medina Tanco (1997a). The results depend, of 
course, on the model used to describe the large scale galactic 
magnetic field. This is certainly a largely unexplored area. 
However, we expect that a rough description, satisfactory for the 
present treatment, can be attained by the models of Stanev (1997). 
We adopt the same two extreme combinations of Stanev (1997)
(see also Sofue et al. 1986, Beck et al, 1996):  (1) a 
bisymmetric GMF model with field reversals and odd parity (BSS-A) 
and (2) an axisymmetric GMF model without reversals and with even 
parity (ASS-S). The effects of a small $B_{z}=0.1$ $\mu$G component are also 
studied in each case.

Table 1, adapted from Hayashida et al. (1996), lists the proposed 
clusters of events. $\Delta t_{arr}$, is the arrival time delay.
The pairs were classified as type A and B, according to the arrival 
order of the highest energy particle. Only type A events, 
where the highest energy particle 
arrives first, can be originated in a bursting source in which the 
particles are simultaneously released. In type B events either the 
source is quiescent  or there is a finite acceleration time involved 
that delays the production of the high energy particle. 

Table 2 summarizes the results for pair 1 under different GMF 
configurations. For an almost simultaneous release of the particles 
at the sources, all the GMF configurations but one, imply that the 
source of pair 1 should lie inside the galactic halo. The maximum 
distance to the source for each one of these GMF models is indicated 
in column 3. Only the  ASS-S model without a $B_{z}$ component allows an 
extragalactic (EG) source. In the later case, a maximum arrival time 
delay $\Delta t_{IGM} \sim 0.6$ yr is left for the intergalactic portion of 
the trajectories of both particles. 

The arrival time delay between a proton of energy E and a photon can 
be estimated as: 

\begin{equation}
t_{p\gamma} \sim 9 \times 10^{4} 
                   \times \left( \frac{B}{10^{-9} G}  \right)^{2}
                   \times \left( \frac{D}{30 Mpc}     \right)^{2}
                   \times \left( \frac{E}{10^{20} eV} \right)^{-2}
                   \times \left( \frac{L_{c}}{1 Mpc}  \right)
                   \mbox{\hspace{1cm} [yr]}
\end{equation}

\noindent (c.f. , Waxman and Coppi, 1996), where $B$, $L_{c}$ and $D$ are 
the intensity of the IGMF, its correlation length and the distance to the 
source respectively.  If the correlation length is known, Eq. (1) can be 
used to estimate maximum IGMF for a given $D$ and $\Delta t_{arr}$
between two protons. Two fiducial distances have 
been selected for quantification purposes: $D=3$ and $D=30 Mpc$. 
The maximum values of the IGMF for these distances are listed in Table 2 
for $L_{c} = 1$ Mpc (Kronberg 1994, 1996). These are
the constraints set upon the intergalactic propagation region and UHECR 
bursting sources by the observed pair 1, after considering the propagation 
of the particles through GMF. However, another constraint must be 
satisfied: the angle between the momenta of the particles 
arriving at the border of the halo from the IGM should be equal to the 
calculated $\theta_{HALO}$ in Table 2.

To this end, numerical simulations (Medina Tanco et al 1997b) were 
carried out emulating the intergalactic propagation of the components 
of pair 1. $L_{c} \sim 1$ Mpc is assumed, while IGMF values and distances to the 
sources are those of Table 2. Protons are injected at the sources with 
an $E^{-2}$ spectrum and the energy losses included are redshift, pair 
production and photo-pion production (Berezinsky and 
Grigor'eva 1988). The resulting distribution functions for the 
relative time delay and the arrival angle between the proton components 
of pair 1 are shown in figures 1 and 2. The 
average time delay between both protons, as given by the simulations 
(figure 1), 
is consistent with equation (1), although there is a considerable 
dispersion. Furthermore, figure 2 shows that a 
$\theta_{HALO} = 2^{o}$ 
separation, as inferred for pair 1 at the border of the halo, is at 
the wing of the distribution. Therefore, if a bursting source were 
responsible for this pair, a very low probability event was observed 
indeed.

Pairs 2 and 3 are type B events. This means that a point source 
cannot have emitted the UHECR simultaneously. Therefore, if the 
point source hypothesis is to be maintained, we must assume either 
that the source is quiescent or, if bursting, that a finite 
acceleration time is involved which delays the emission of the 
high energy component. In this case, the sum of the arrival time 
delay and the time delay due to propagation through the GMF and IGMF, 
is a lower limit to $\tau_{s}$, the lifetime of the source. 
Again, the galactic and intergalactic trajectories must verify the 
matching of $\theta_{HALO}$ at the galactic halo border. 
It is found that 
$\theta_{HALO}$(pair 2) $\sim 2^{o}$ and 
$\theta_{HALO}$(pair 3) $\sim 2^{o}-5.5^{o}$, 
depending on the GMF model adopted. 
Numerical simulations for the IGM propagation of the proton 
components of pairs 2 and 3 are also shown in figures 1 and 2. 
$D = 30$ Mpc, and $B_{IGM} = 10^{-12}$ and $10^{-9}$ Gauss, 
were used. 
The lower value of the IGMF is the one imposed by  a bursting 
pair 1, and the second is the current upper limit for the IGMF. 
It can be seen from figures 1 and 2 that, as for pair 1, 
a $10^{-12}$ Gauss IGMF leads to a very low probability 
for an event with $\theta_{HALO}$ on the order of a few degrees. 
Taking into account the galactic propagation, the lower limits for 
the lifetime of single sources for pairs 2 and 3, with 
$B_{IGM} \sim 10^{-12}$ Gauss, are $\sim 10$ and $\sim 100$ yr 
respectively. On the other hand, from the point of view of 
$\theta_{HALO}$, a consistent picture can be obtained for a 
higher value of the IGMF, say near $10^{-9}$ Gauss. However, 
$\tau_{s} > 10^{5}$ yr and so a single source should be quiescent 
and, probably, extended perhaps enclosing more than one galaxy in 
order to confine $\sim 10^{20}$ eV particles. 

The previous results seem to point to a chance clustering of the 
three pairs of events, despite 
the chance probability for the pairs quoted by 
Hayashida et al. (1996) being only 2.9\%. We note, however, that 
this chance probability was derived under the assumption that 
the arrival direction distribution is uniform over the sky. 
This is arguable. Several classes of potential extragalactic 
sources have been proposed (e.g., Kewley et al 1996, Protheroe 
and Johnson 1996, Biermann et al 1996, Halzen 1997), and these 
are not uniformly distributed over the sky. The inhomogeneity 
of the source's distribution should be more noticeable because 
the interaction of UHECR with the cosmic microwave background 
(CMB) imposes an upper limit 
$D_{max} \sim 10^{2}$ Mpc. Even if the actual sources are unknown, 
we can naively assume that they follow the distribution of galaxies 
(i.e., luminous matter)
in the nearby universe.  This is compatible  
with isolated galaxies, interacting galaxies, galactic 
bowshocks in high density IGMs and extended sources in turbulent 
IGMs powered by concentrations of galaxies. 

Except for some few observational  determinations and 
upper limits (e.g., Arp  1988,  Kim et al. 1989, Kronberg 1994) or 
numerical simulations of cosmological structure formation (Biermann 
1996 and references there in) we know very little about the IGMF. These 
constraints, however, point to an IGMF structure that 
follows the distribution of matter (galaxies). 
Therefore, a high degree of inhomogeneity  can be expected, 
with relatively high values of $B_{IGM}$  over small regions ($\sim 1$ Mpc) 
of  high matter density (c.f., Arp 1988, Kim et al. 1989), 
pervading vast low density/low $B_{IGM}$ regions with 
$B_{IGM} < 10^{-9}$ G. 

Following Medina Tanco et al (1997b), it is assumed that 
the UHECR are protons, and that their sources are extragalactic and 
hosted by, or associated with, normal galaxies. It is further assumed 
that the magnetic field scales as $n_{gal}^{2/3}$, where $n_{gal}$ is 
the local density of galaxies as derived from the CfA redshift 
catalogue (Huchra et al 1995). The IGMF is considered as organised 
in cells of size $L_{c}$ of homogeneous field, such that the orientation 
of $B_{IGM}$ between adjacent cells is uncorrelated. 
$L_{c}$ relates to the IGMF through the expression: 
$L_{c}(r) \propto [B_{IGM}(r)]^{-2}$, 
and the normalisation condition $L_{c} \sim 1$ Mpc for 
$B_{IGM} \sim 10^{-9}$ G is adopted. UHECR protons are injected at 
the galaxies with an energy spectrum $\propto E^{-2}$ 
and propagated non-diffusively through the above scenario, while  
loosing energy via redshift, pair production and photomeson production 
(Berezinsky and Grigor'eva 1988).

The results are displayed in the form of all-sky UHECR images of the 
celestial sphere for galaxies located at $20 < D < 50$ Mpc (Figure 3.a) 
and $50 < D < 200$ Mpc (Figure 3.b) for arriving protons with 
$E > 4 \times 10^{19}$ eV. 
These surfaces should be representative of the 
arrival probability of UHECR at the Earth position in the Galaxy. 
The curved lines bound 
the region of the sky where AGASA is believed to be sensitive 
(Uchihori 1996). 

We can see that the arrival probability is by no means isotropic. 
Furthermore, pair 2 is on top of a maximum of the arrival probability 
for sources located between $20$ and $50$ Mpc, while pair 1 is also 
located on a high arrival probability region for sources at more than 
$50$ Mpc. This is in contrast with the chance probability estimated 
by Hayashida et al. (1996), and points to either different uncorrelated 
sources of the components of each pair, or to very extended quiescent 
sources involving several galaxies. We also note that the sensitivity 
of AGASA is rather low in the vicinity of pairs 1 and 2. Therefore, an 
instrument with more uniform coverage (like the proposed Auger project) 
should probably detect an extended region of excess UHECR flux at the 
position of the pairs.

The third pair comes from a region of space where no large clustering 
of galaxies exist up to the depths considered. As the components 
cannot have originated simultaneously at the same extragalactic 
source because of galactic propagation constraints, they must have 
come from isolated sources. This seems to indicate that very large 
agglomerates of galaxies, large enough to give a signature in 
figure 3, are not needed in order to accelerate UHECR.

\section{Conclusions}

The constraints deduced from the propagation of the components of the 
pairs of UHECR events proposed by Hayashida et al (1996) through the 
galactic and intergalactic medium have been analysed. 

In the case of pair 1, the low value of the IGMF, imposed by the 
arrival time delay between the protons, is inconsistent with the 
deflection angle between the momenta of the particles at the border 
of the halo, inferred from their galactic propagation.  This makes a 
single, bursting source very unlikely. 

If the components of pairs 2 and 3 originate in common sources, then 
the lifetimes of the sources are probably larger than few times 
$10^{5}$ yr and, therefore, extended. This picture is consistent with 
an IGMF value not much smaller than the presently accepted upper limit
of $10^{-9}$ Gauss (Kronberg 1996) and a distance to the sources of 
$\sim 30$ Mpc. In fact, the actual distribution of galaxies 
(Huchra et al, 1995) presents a local maximum at about that distance 
in the direction of pair 2. Furthermore, our simulations point 
to a maximum in the arrival distribution of UHECR at exactly the same 
position of pair 2, when sources between $20$ and $50$ Mpc are considered. 
Pair 1 is also located inside a maximum of the arrival distribution, 
favouring chance pairing between the components.


This work was done with the partial support from the Brazilian agency FAPESP.

\newpage







\noindent
{\bf Figure Captions}

\bigskip

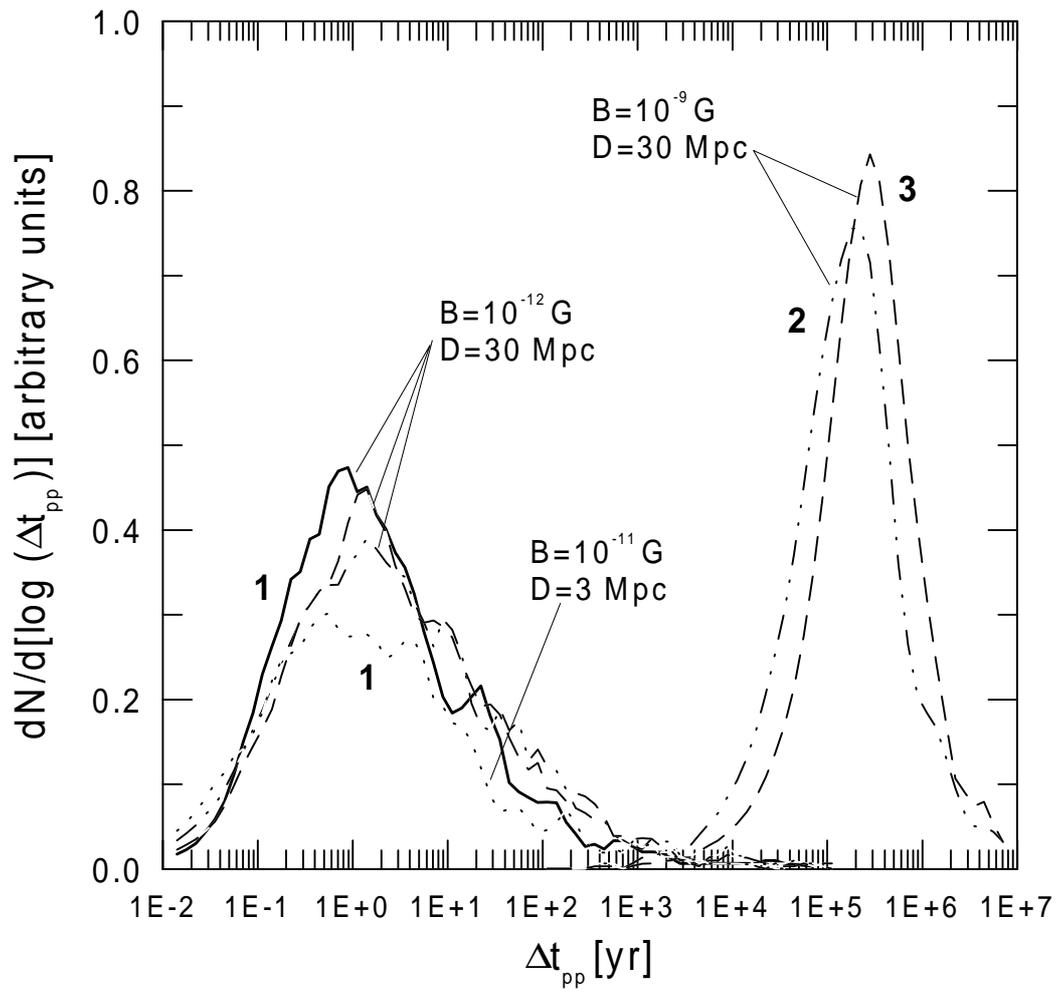
Figure 1:  Distribution function of arrival time delays between the 
observed pair of protons in clusters 1, 2 and 3 due to propagation in 
the IGMF alone (i.e., at the external border of the galactic). halo). 
The simulations for pair 1 correspond to the two fiducial scenarios of 
Table 2, and match the constrain in time delay imposed by the galactic 
portion of the tracks: $B_{IGM}=10^{-11}$ G and $D=3$ Mpc (dotted line) 
and $B_{IGM} = 10^{-12}$ G and $D = 30$ Mpc (continuous line). For 
pairs 2 (broken-dotted line) and 3 (broken line) two possible scenarios 
are explored: $D = 30$ Mpc and $B_{IGM}=10^{-12}$ G and 
$B_{IGM}=10^{-9}$ G.

\bigskip

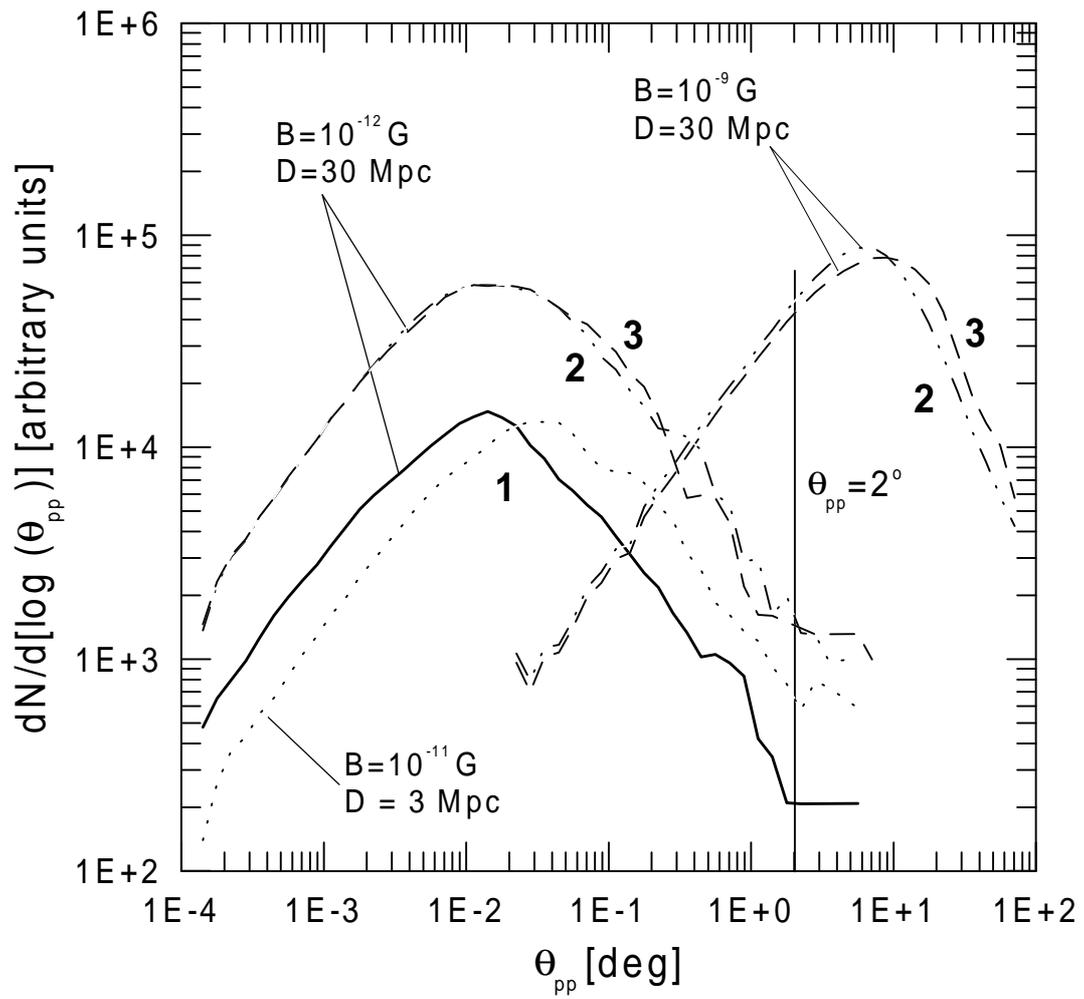
Figure 2: Distribution function of  the angle between the momenta of 
the observed particles in each pair, at their arrival at the external 
border of the galactic halo after propagation through the IGMF. The 
conditions are the same as in figure 1. Also indicated is a separation 
angle of $2^{o}$ typically obtained from the calculations of galactic 
propagation for all the three pairs.

\bigskip

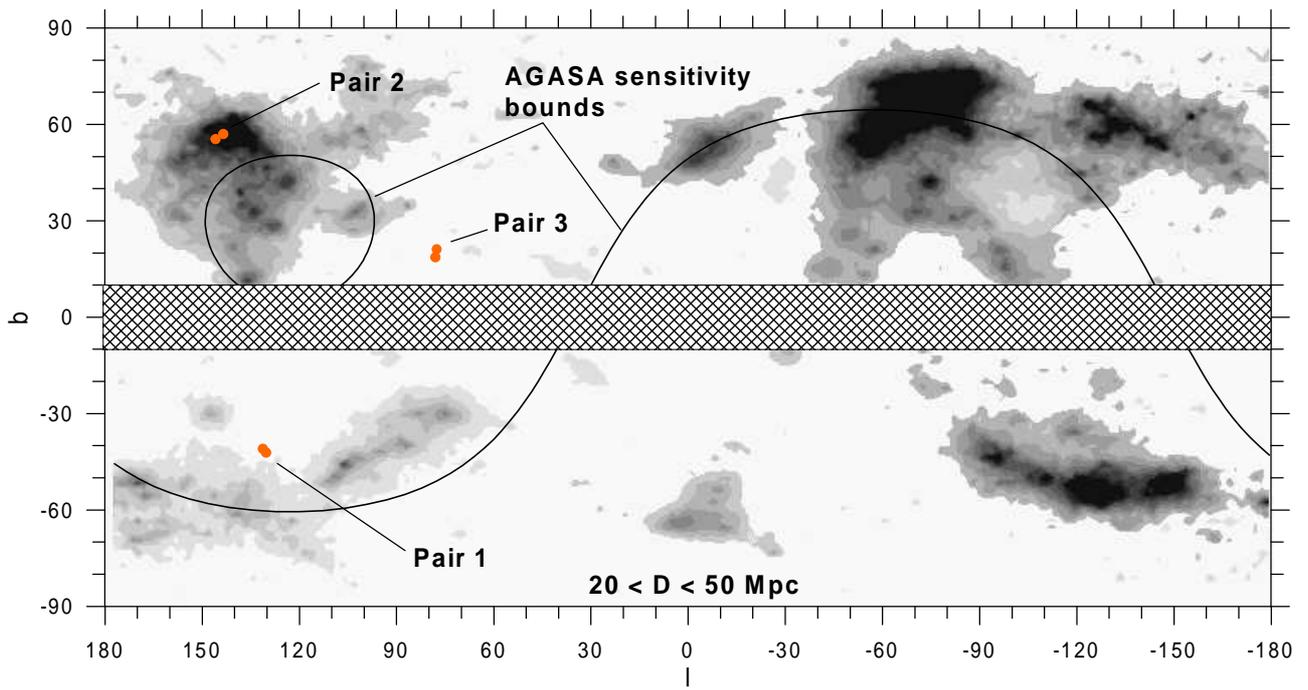
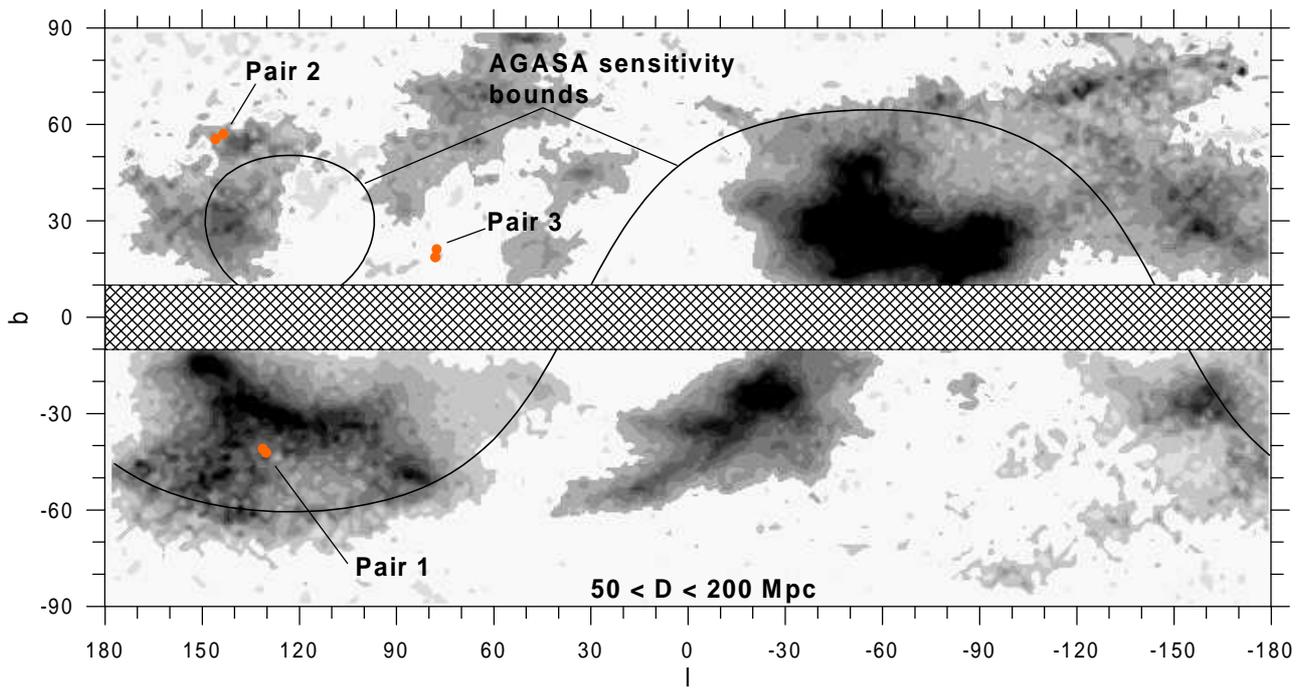
Figure 3:  Arrival distribution of UHECR simulated under the assumption 
that the luminous matter in the nearby Universe tracks the distribution 
of the sources of UHECR, and modulates the intensity of the IGMF (see 
text for details). Redshift, pair production and photo-pion production 
losses are included. The arrival distributions due to sources at two 
different depths are shown: (a) $0 < D < 50$ Mpc, (b) $50 < D < 200$ Mpc. 
Pairs 1 and 2 lie on top of regions of high arrival probability 
strengthening the possibility of chance pairing. The solid lines 
bound the region of the sky actually seen by AGASA. Sensitivity is 
poor near these lines


\begin{table}
\begin{center}
\caption{Possible Clusters of UHECRs observed by AGASA 
experiment (adapted from Hayashida et al, 1996)}
\begin{tabular}{ccccccc}
\tableline
\tableline
Pair No.      & 
Date         &
 $\Delta t_{arr}$ [yr]    & 
Energy [eV]   &
Type   &
$l_{gal}$  &
$b_{gal}$ \\
\tableline
1       & 93/12/03 &   1.90  & 210 &    A    & 131.2 & -41.1 \\
\nodata & 95/10/29 & \nodata &  51 & \nodata & 130.2 & -42.3 \\
\tableline
2       & 92/08/01 &   2.49  &  55 &    B    & 143.5 &  56.9 \\
\nodata & 95/01/26 & \nodata &	78 & \nodata & 145.8 &  55.3 \\
\tableline
3       & 91/04/20 &   3.21  &  43 &    B    &  77.9 &  18.6 \\
\nodata & 94/07/06 & \nodata & 110 & \nodata &  77.6 &  21.1 \\
\tableline
\tableline
\end{tabular}
\end{center}
\end{table}

\begin{table}
\begin{center}
\caption{Pair1: Constraints from galactic and intergalactic propagation.}
\begin{tabular}{ccccccc}
\tableline
\tableline
$B_{gal}$      & 
$B_{z}$          &
 $D_{max}$ [kpc]     & 
$\Delta t_{IGM}$ [yr] &
$\theta t_{HALO}$ [$^{o}$]   &
$B_{IGM}^{max}(3 Mpc)$   &
$B_{IGM}^{max}(30 Mpc)$ \\
\tableline
ASS-S    &   $=0$   &  EG  &    0.6    &  2    & $10^{-11}$ & $10^{-12}$\\
\tableline
\nodata  &  $\ne0$  &   7  &  \nodata  &  4.5  &  \nodata   & \nodata   \\
\tableline
BSS-A    &   $=0$   &   8  &  \nodata  &  2    &  \nodata   & \nodata   \\
\tableline
\nodata  &  $\ne0$  &  13  &  \nodata  &  0.5  &  \nodata   & \nodata   \\
\tableline
\tableline
\end{tabular}
\end{center}
\end{table}

\end{document}